\documentclass{webofc}

\usepackage[varg]{txfonts}   
\usepackage{hyperref}
\usepackage{url}
\hypersetup{colorlinks=true,citecolor=blue,urlcolor=blue,linkcolor=blue}
\begin{document}
\title{Strangeness in Astrophysics}
%
%

\author{\firstname{Laura} \lastname{Tolos}\inst{1,2,3}\fnsep\thanks{\email{tolos@ice.csic.es}} }

\institute{Institute of Space Sciences (ICE, CSIC), Campus UAB, Carrer de Can Magrans, 08193, Barcelona, Spain
\and
      Institut d'Estudis Espacials de Catalunya (IEEC), 08860 Castelldefels (Barcelona), Spain
\and
Frankfurt Institute for Advanced Studies,  University of Frankfurt, Ruth-Moufang-Str. 1, 60438 Frankfurt am Main, Germany
          }

\abstract{In this contribution the role of strangeness in astrophysics is discussed and, more precisely, strange hadronic matter in the interior of neutron stars. A special attention is payed to certain phenomena involving strange hadronic matter, such as the hyperon puzzle, kaon condensation and the thermal behaviour of hyperons in neutron star mergers.}
\maketitle

\section{Introduction}
\label{sec:eos-strange}
Compact stars and, more precisely, neutron stars (NSs) are an excellent laboratory to study matter under extreme conditions of density, isospin asymmetry and temperature \cite{Tolos:2020aln,Burgio:2021vgk}. NSs result from core-collapse supernovae and are in hydrostatic equilibrium with the gravitational collapse counterbalanced by the pressure inside NSs.  NSs usually have masses around 1-2~M$_{\odot}$ and radii about 10-12 km, with an onion-like configuration. The largest region of an NS is the inner region, whose composition is not known.
Thus, several hypothesis have been put forward to describe the NS interior. 

Among others, hadronic matter made of strange baryons (also called hyperons) and/or strange mesons (mainly antikaons) has been postulated as possible scenarios in the inner core of NSs. In this contribution we aim at reviewing some specific phenomena related to the presence of strange hadronic matter inside NSs, such as the hyperon puzzle, the possible existence of kaon condensation as well as the thermal behaviour of hyperons as a possible indicator of their presence in NS mergers.

\section{The hyperon puzzle}

Hyperons are baryons with one or more strange quarks, such as $\Lambda$, $\Sigma$ and $\Xi$.  Since the seminal work of Ref.~\cite{1960SvA.....4..187A}, hyperons have been thoroughly studied in the inner core of NS (for recent reviews see \cite{Tolos:2020aln,Burgio:2021vgk,Vidana:2018bdi,Chatterjee:2015pua}). NSs are charged neutral objects equilibrated by weak interactions involving the different species inside the core ($\beta$-equilibrium). The equilibrium can be expressed in terms of the chemical potentials of the different components. In particular, hyperons are expected to appear in the interior of NSs at densities of $\approx 2$-$3 \rho_0$, as the nucleon chemical potential could be so large at these densities that is energetically more favourable to have hyperons rather than highly energetic nucleons at the top of the Fermi seas. As a consequence, matter becomes softer when only nucleons are present as the system relieves Fermi pressure. Hence, the less pressure there is inside an NS, the less mass it can sustain. The softening of the equation of state (EoS) of matter could then lead to maximum masses not compatible with  $2M_{\odot}$ measurements, such as the masses of the PSR J1614-2230 \cite{Demorest:2010bx,Fonseca:2016tux}, PSR J0348+0432 \cite{Antoniadis:2013pzd} and PSR J0740+6620 \cite{2019NatAs.tmp..439C}. This fact is usually denoted as {\it the hyperon puzzle}. There is, however, no solution yet to this puzzle, although several answers have been put forward in order to have hyperons in the interior of $2M_{\odot}$ NSs. 

Among them, one solution considers stiff hyperon-nucleon and hyperon-hyperon interactions so as to compensate for the softening of the EoS (see, for example, \cite{Bednarek:2011gd,Weissenborn:2011ut,Oertel:2014qza,Maslov:2015msa}).  Another way for stiffening the EoS relies on the three-body forces \cite{Haidenbauer:2016vfq,Logoteta:2019utx,Gerstung:2020ktv,Takatsuka2002,Takatsuka:2004ch,Vidana:2010ip,Yamamoto:2013ada,Yamamoto:2014jga,Lonardoni:2014bwa}.
Another solution pushes hyperons to larger densities by the occurrence of new species, such as $\Delta$ baryons \cite{Schurhoff:2010ph,Drago:2014oja,Ribes:2019kno} or a kaon condensate (as discussed in the following section). Also, solutions based on non-hadronic degrees of freedom have been considered, such as an early phase transition to quark matter below the hyperon onset (see Refs.~\cite{Weissenborn:2011qu,Bonanno:2011ch,Klahn:2013kga,Lastowiecki:2011hh,Zdunik:2012dj} for recent papers).  And, finally, modified gravity theories allowing for hyperons inside $2M_{\odot}$ stars   have been also described in the literature \cite{Astashenok:2014pua}.

\section{Kaon condensation in neutron stars}
\label{sec:kaoncond}

Another possibility for strange hadronic matter is to consider strange mesons, such as antikaons in the core of NSs.  As mentioned before, the composition of matter in NSs is determined by demanding $\beta$-equilibrium. In the particular case of matter made of neutrons, protons and electrons, the chemical potential of the electron substantially increases with density. Thus, antikaons could be produced if their effective mass gets smaller than the electron chemical potential at a given density inside NSs. In that case, antikaons would appear forming a condensate, hence leading to the phenomenon of {\it kaon condensation}. 

Since the pioneering work of Ref.~\cite{Kaplan:1986yq}, this phenomenon has been widely discussed.  The question is whether the mass of antikaons could be largely modified in nuclear matter. Whereas this is the case for some phenomenological models  (see the recent results in Refs.~\cite{Gupta:2013sna,Thapa:2021kfo,Malik:2021nas,Muto:2021jms}), microscopic unitarized schemes do not obtain such large modifications in the antikaon mass (see, for example, Refs.~\cite{Lutz:1997wt,Ramos:1999ku,Tolos:2000fj,Tolos:2002ud,Tolos:2006ny,Tolos:2008di,Cabrera:2014lca,Lutz:2007bh}).

\section{Thermal behaviour of hyperons in neutron star mergers}
\label{sec:mergers}

In spite of the previous scenarios where strange hadronic matter might be present inside an NS, there is still the open question whether there is a clear signal of the presence of strange hadronic matter, and in particular hyperons, in NSs. For example, by comparing the mass-radius relations coming from EoSs with only nucleons or with nucleons and hyperons,  it is not straightforward to tell from a measured mass-radius relation if the underlying EoS contains hyperons. Therefore, it is very much welcome to identify NS features that can be clearly linked to the appearance of hyperons in NS. 

Recently, the {\it thermal behavior of hyperons} has been proposed as a potential indicator for the presence of hyperons in NS mergers \cite{Blacker:2023opp}.  The approach is based on the fact that the hyperonic EoSs  produce lower average thermal indeces in matter than the  nucleonic ones \cite{Kochankovski:2022rid,Raduta:2022elz}. The influence of the different thermal treatment on the dominant postmerger GW frequency is then systematically investigated using numerical simulations  \cite{Blacker:2023opp}.

\begin{figure}
\centering{\includegraphics[width=0.495\linewidth]{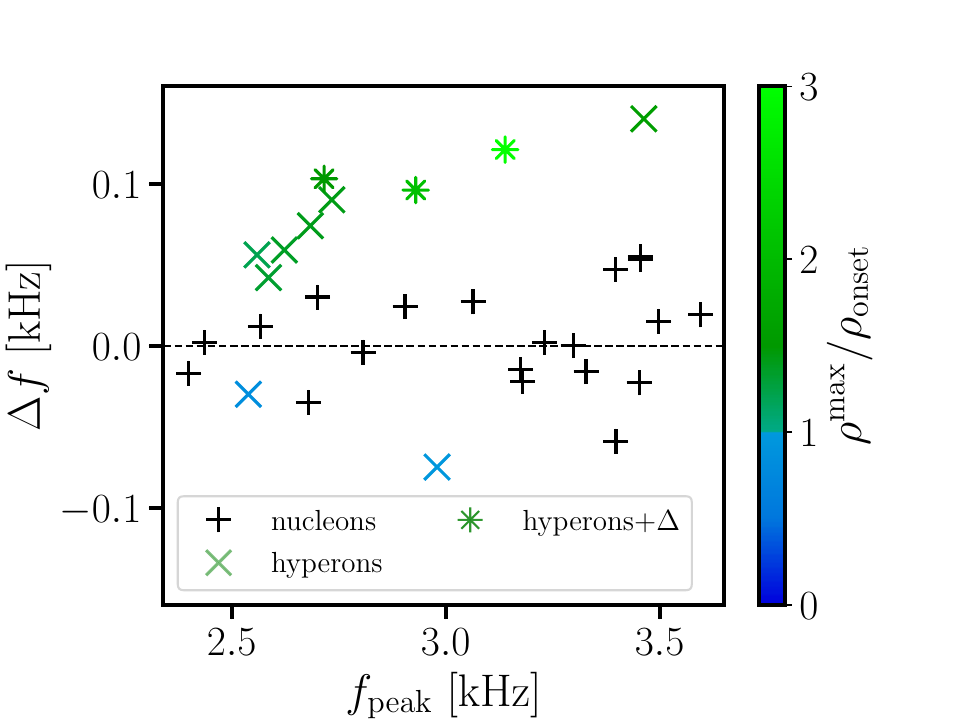}
\includegraphics[width=0.495\linewidth]{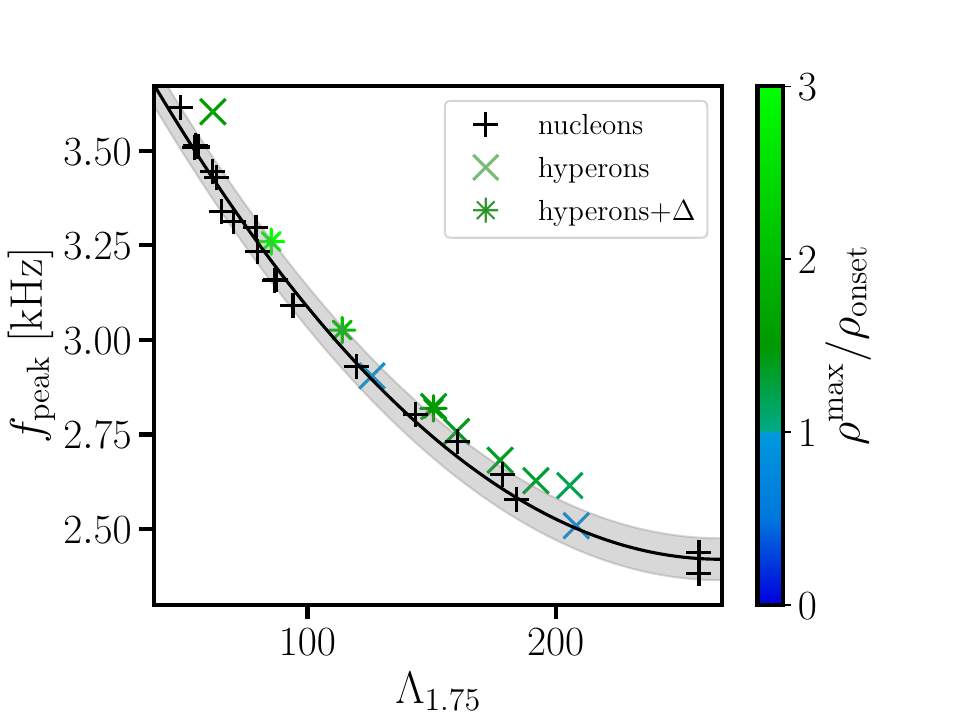}}
\caption{Left plot: $\Delta f=f_\mathrm{peak}-f_\mathrm{peak}^{1.75}$ as a function of $f_\mathrm{peak}$. Right plot: $f_\mathrm{peak}$ of 1.4-1.4~$M_\odot$ mergers as function of tidal deformability of a 1.75~$M_\odot$ NS.  Black symbols depict purely nucleonic models, whereas crosses display hyperonic models and asterisks refer to models which include $\Delta$s. The coloring shows the ratio between the maximum rest-mass density in the postmerger remnant and the hyperon onset rest-mass density at zero temperature. Black curve on the right plot shows least-squares fit to purely nucleonic models, while grey band displays maximum residual of purely nucleonic models from the fit. Plots taken from \cite{Blacker:2023opp}.} 
\label{fig:df}
\end{figure}

First,  this frequency is obtained by running simulations using the full temperature- and composition-dependent nucleonic and hyperonic EoSs. Then, in another set of simulations, all EoSs at zero temperature in neutrinoless $\beta$-equilibrium are used and supplemented with the ideal-gas treatment of thermal pressure with $\Gamma_\mathrm{th}=1.75$, mimicking the thermal behavior of purely nucleonic matter. And, finally, the dominant postmerger GW frequencies of the full EoS models, $f_\mathrm{peak}$, with the frequencies $f_\mathrm{peak}^{1.75}$ obtained from the $\Gamma_\mathrm{th}=1.75$ calculations are compared, so that  $\Delta f \equiv f_\mathrm{peak}-f_\mathrm{peak}^{1.75}$ measures the deviation from an idealized nucleonic thermal behavior. The results are shown in the l.h.s of Fig.~\ref{fig:df}, distinguishing purely nucleonic and hyperonic models.  A characteristic increase of the dominant postmerger gravitational-wave frequency by up to $\sim150$~Hz is observed with the hyperonic EoS models (with a sufficient amount of hyperons) compared to purely nucleonic ones, thus linking this effect to the occurrence of hyperons in NSs. 

Also, regarding the identification of hyperons by a frequency shift within a more concrete detection scenario, it is interesting to relate $f_\mathrm{peak}$ from the simulations with the fully temperature-dependent EoSs as function of the tidal deformability of 1.75~$M_\odot$ NSs, as done in the r.h.s of Fig.~\ref{fig:df}.The dominant postmerger frequency of hyperonic models is characteristically increased compared to the nucleonic models, so that  the presence of  hyperons may be deduced by an increased postmerger frequency that  is not compatible with a nucleonic EoS. 

Some caveats from the previous analysis need, however, to be mentioned, such as the necessity of enough statistical power in GW measurements, the dependence on the abundance of hyperons and hyperon threshold, or the fact that other exotic degrees of freedom might lead to a similar frequency shift. 

\section{Conclusions}
\label{sec:conclusions}
In this contribution  certain phenomena associated to the occurrence of strange hadronic matter in the inner core of NSs have been presented. More precisely, we have reviewed the hyperon puzzle and the possible solutions as well as the potential existence of a kaon condensed phase in NSs. Furthermore, the thermal behaviour of hyperons in NS mergers has been discussed as a feasible indicator of strangeness in the inner core of NSs.

\section*{Acknowledgments}

L.T. thanks H. Kochankovski and A. Ramos for the careful reading of this contribution. She also acknowledges support from CEX2020-001058-M (Unidad de Excelencia ``Mar\'{\i}a de Maeztu") and PID2022-139427NB-I00 financed by the Spanish MCIN/AEI/10.13039/501100011033/FEDER,UE as well as from the Generalitat de Catalunya under contract 2021 SGR 171,  by  the EU STRONG-2020 project, under the program  H2020-INFRAIA-2018-1 grant agreement no. 824093, and by the CRC-TR 211 'Strong-interaction matter under extreme conditions'- project Nr. 315477589 - TRR 211. 

%

\bibliography{tolos_SQM2024.bib}

\begin{thebibliography}{47}

\bibitem{Tolos:2020aln}
L.~Tolos, L.~Fabbietti, {Strangeness in Nuclei and Neutron Stars}, Prog. Part.
  Nucl. Phys. \textbf{112}, 103770 (2020), \texttt{2002.09223}.
  \doiwoc{10.1016/j.ppnp.2020.103770}

\bibitem{Burgio:2021vgk}
G.F. Burgio, H.J. Schulze, I.~Vidana, J.B. Wei, {Neutron stars and the nuclear
  equation of state}, Prog. Part. Nucl. Phys. \textbf{120}, 103879 (2021),
  \texttt{2105.03747}. \doiwoc{10.1016/j.ppnp.2021.103879}

\bibitem{1960SvA.....4..187A}
V.A. {Ambartsumyan}, G.S. {Saakyan}, {The Degenerate Superdense Gas of
  Elementary Particles}, Soviet Astronomy \textbf{4}, 187 (1960).

\bibitem{Vidana:2018bdi}
I.~Vida\~na, {Hyperons: the strange ingredients of the nuclear equation of
  state}, Proc. Roy. Soc. Lond. \textbf{A474}, 0145 (2018),
  \texttt{1803.00504}. \doiwoc{10.1098/rspa.2018.0145}

\bibitem{Chatterjee:2015pua}
D.~Chatterjee, I.~Vida\~na, {Do hyperons exist in the interior of neutron
  stars?}, Eur. Phys. J. A \textbf{52}, 29 (2016), \texttt{1510.06306}.
  \doiwoc{10.1140/epja/i2016-16029-x}

\bibitem{Demorest:2010bx}
P.~Demorest, T.~Pennucci, S.~Ransom, M.~Roberts, J.~Hessels, {Shapiro Delay
  Measurement of A Two Solar Mass Neutron Star}, Nature \textbf{467}, 1081
  (2010), \texttt{1010.5788}. \doiwoc{10.1038/nature09466}

\bibitem{Fonseca:2016tux}
E.~Fonseca et~al., {The NANOGrav Nine-year Data Set: Mass and Geometric
  Measurements of Binary Millisecond Pulsars}, Astrophys. J. \textbf{832}, 167
  (2016), \texttt{1603.00545}. \doiwoc{10.3847/0004-637X/832/2/167}

\bibitem{Antoniadis:2013pzd}
J.~Antoniadis et~al., {A Massive Pulsar in a Compact Relativistic Binary},
  Science \textbf{340}, 6131 (2013), \texttt{1304.6875}.
  \doiwoc{10.1126/science.1233232}

\bibitem{2019NatAs.tmp..439C}
H.T. Cromartie et~al., {Relativistic Shapiro delay measurements of an extremely
  massive millisecond pulsar}, Nature Astronomy p. 439 (2019),
  \texttt{1904.06759}. \doiwoc{10.1038/s41550-019-0880-2}

\bibitem{Bednarek:2011gd}
I.~Bednarek, P.~Haensel, J.L. Zdunik, M.~Bejger, R.~Manka, {Hyperons in
  neutron-star cores and two-solar-mass pulsar}, Astron. Astrophys.
  \textbf{543}, A157 (2012), \texttt{1111.6942}.
  \doiwoc{10.1051/0004-6361/201118560}

\bibitem{Weissenborn:2011ut}
S.~Weissenborn, D.~Chatterjee, J.~Schaffner-Bielich, {Hyperons and massive
  neutron stars: vector repulsion and SU(3) symmetry}, Phys. Rev. \textbf{C85},
  065802 (2012), [Erratum: Phys. Rev.C90,no.1,019904(2014)],
  \texttt{1112.0234}. \doiwoc{10.1103/PhysRevC.85.065802,
  10.1103/PhysRevC.90.019904}

\bibitem{Oertel:2014qza}
M.~Oertel, C.~Providência, F.~Gulminelli, A.R. Raduta, {Hyperons in neutron
  star matter within relativistic mean-field models}, J. Phys. \textbf{G42},
  075202 (2015), \texttt{1412.4545}. \doiwoc{10.1088/0954-3899/42/7/075202}

\bibitem{Maslov:2015msa}
K.A. Maslov, E.E. Kolomeitsev, D.N. Voskresensky, {Solution of the Hyperon
  Puzzle within a Relativistic Mean-Field Model}, Phys. Lett. \textbf{B748},
  369 (2015), \texttt{1504.02915}. \doiwoc{10.1016/j.physletb.2015.07.032}

\bibitem{Haidenbauer:2016vfq}
J.~Haidenbauer, U.G. Mei{\ss}ner, N.~Kaiser, W.~Weise, {Lambda-nuclear
  interactions and hyperon puzzle in neutron stars}, Eur. Phys. J.
  \textbf{A53}, 121 (2017), \texttt{1612.03758}.
  \doiwoc{10.1140/epja/i2017-12316-4}

\bibitem{Logoteta:2019utx}
D.~Logoteta, I.~Vida\~na, I.~Bombaci, {Impact of chiral hyperonic three-body
  forces on neutron stars}, Eur. Phys. J. \textbf{A55}, 207 (2019),
  \texttt{1906.11722}. \doiwoc{10.1140/epja/i2019-12909-9}

\bibitem{Gerstung:2020ktv}
D.~Gerstung, N.~Kaiser, W.~Weise, {Hyperon\textendash{}nucleon three-body
  forces and strangeness in neutron stars}, Eur. Phys. J. A \textbf{56}, 175
  (2020), \texttt{2001.10563}. \doiwoc{10.1140/epja/s10050-020-00180-2}

\bibitem{Takatsuka2002}
T.~Takatsuka, S.~Nishizaki, Y.~Yamamoto, Necessity of extra repulsion in
  hypernuclear systems: Suggestion from neutron stars, The European Physical
  Journal A - Hadrons and Nuclei \textbf{13}, 213 (2002).
  \doiwoc{10.1140/epja1339-35}

\bibitem{Takatsuka:2004ch}
T.~Takatsuka, {Hyperon-mixed neutron stars}, Prog. Theor. Phys. Suppl.
  \textbf{156}, 84 (2004). \doiwoc{10.1143/PTPS.156.84}

\bibitem{Vidana:2010ip}
I.~Vida\~na, D.~Logoteta, C.~Providencia, A.~Polls, I.~Bombaci, {Estimation of
  the effect of hyperonic three-body forces on the maximum mass of neutron
  stars}, EPL \textbf{94}, 11002 (2011), \texttt{1006.5660}.
  \doiwoc{10.1209/0295-5075/94/11002}

\bibitem{Yamamoto:2013ada}
Y.~Yamamoto, T.~Furumoto, N.~Yasutake, T.A. Rijken, {Multi-pomeron repulsion
  and the Neutron-star mass}, Phys. Rev. \textbf{C88}, 022801 (2013),
  \texttt{1308.2130}. \doiwoc{10.1103/PhysRevC.88.022801}

\bibitem{Yamamoto:2014jga}
Y.~Yamamoto, T.~Furumoto, N.~Yasutake, T.A. Rijken, {Hyperon mixing and
  universal many-body repulsion in neutron stars}, Phys. Rev. \textbf{C90},
  045805 (2014), \texttt{1406.4332}. \doiwoc{10.1103/PhysRevC.90.045805}

\bibitem{Lonardoni:2014bwa}
D.~Lonardoni, A.~Lovato, S.~Gandolfi, F.~Pederiva, {Hyperon Puzzle: Hints from
  Quantum Monte Carlo Calculations}, Phys. Rev. Lett. \textbf{114}, 092301
  (2015), \texttt{1407.4448}. \doiwoc{10.1103/PhysRevLett.114.092301}

\bibitem{Schurhoff:2010ph}
T.~Schuerhoff, S.~Schramm, V.~Dexheimer, {Neutron stars with small radii -- the
  role of delta resonances}, Astrophys. J. \textbf{724}, L74 (2010),
  \texttt{1008.0957}. \doiwoc{10.1088/2041-8205/724/1/L74}

\bibitem{Drago:2014oja}
A.~Drago, A.~Lavagno, G.~Pagliara, D.~Pigato, {Early appearance of $\Delta$
  isobars in neutron stars}, Phys. Rev. \textbf{C90}, 065809 (2014),
  \texttt{1407.2843}. \doiwoc{10.1103/PhysRevC.90.065809}

\bibitem{Ribes:2019kno}
P.~Ribes, A.~Ramos, L.~Tolos, C.~Gonzalez-Boquera, M.~Centelles, {Interplay
  between $\Delta$ Particles and Hyperons in Neutron Stars}, Astrophys. J.
  \textbf{883}, 168 (2019), \texttt{1907.08583}.
  \doiwoc{10.3847/1538-4357/ab3a93}

\bibitem{Weissenborn:2011qu}
S.~Weissenborn, I.~Sagert, G.~Pagliara, M.~Hempel, J.~Schaffner-Bielich, {Quark
  Matter In Massive Neutron Stars}, Astrophys. J. \textbf{740}, L14 (2011),
  \texttt{1102.2869}. \doiwoc{10.1088/2041-8205/740/1/L14}

\bibitem{Bonanno:2011ch}
L.~Bonanno, A.~Sedrakian, {Composition and stability of hybrid stars with
  hyperons and quark color-superconductivity}, Astron. Astrophys. \textbf{539},
  A16 (2012), \texttt{1108.0559}. \doiwoc{10.1051/0004-6361/201117832}

\bibitem{Klahn:2013kga}
T.~Klaehn, R.~Lastowiecki, D.B. Blaschke, {Implications of the measurement of
  pulsars with two solar masses for quark matter in compact stars and heavy-ion
  collisions: A Nambu–Jona-Lasinio model case study}, Phys. Rev.
  \textbf{D88}, 085001 (2013), \texttt{1307.6996}.
  \doiwoc{10.1103/PhysRevD.88.085001}

\bibitem{Lastowiecki:2011hh}
R.~Lastowiecki, D.~Blaschke, H.~Grigorian, S.~Typel, {Strangeness in the cores
  of neutron stars}, Acta Phys. Polon. Supp. \textbf{5}, 535 (2012),
  \texttt{1112.6430}. \doiwoc{10.5506/APhysPolBSupp.5.535}

\bibitem{Zdunik:2012dj}
J.L. Zdunik, P.~Haensel, {Maximum mass of neutron stars and strange
  neutron-star cores}, Astron. Astrophys. \textbf{551}, A61 (2013),
  \texttt{1211.1231}. \doiwoc{10.1051/0004-6361/201220697}

\bibitem{Astashenok:2014pua}
A.V. Astashenok, S.~Capozziello, S.D. Odintsov, {Maximal neutron star mass and
  the resolution of the hyperon puzzle in modified gravity}, Phys. Rev.
  \textbf{D89}, 103509 (2014), \texttt{1401.4546}.
  \doiwoc{10.1103/PhysRevD.89.103509}

\bibitem{Kaplan:1986yq}
D.B. Kaplan, A.E. Nelson, {Strange Goings on in Dense Nucleonic Matter}, Phys.
  Lett. \textbf{B175}, 57 (1986). \doiwoc{10.1016/0370-2693(86)90331-X}

\bibitem{Gupta:2013sna}
N.~Gupta, P.~Arumugam, {Impact of hyperons and antikaons in an extended
  relativistic mean-field description of neutron stars}, Phys. Rev. C
  \textbf{88}, 015803 (2013). \doiwoc{10.1103/PhysRevC.88.015803}

\bibitem{Thapa:2021kfo}
V.B. Thapa, M.~Sinha, J.J. Li, A.~Sedrakian, {Massive $\Delta$-resonance
  admixed hypernuclear stars with antikaon condensations}, Phys. Rev. D
  \textbf{103}, 063004 (2021), \texttt{2102.08787}.
  \doiwoc{10.1103/PhysRevD.103.063004}

\bibitem{Malik:2021nas}
T.~Malik, S.~Banik, D.~Bandyopadhyay, {Equation-of-state Table with Hyperon and
  Antikaon for Supernova and Neutron Star Merger}, Astrophys. J. \textbf{910},
  96 (2021), \texttt{2104.00775}. \doiwoc{10.3847/1538-4357/abe860}

\bibitem{Muto:2021jms}
T.~Muto, T.~Maruyama, T.~Tatsumi, {Effects of three-baryon forces on kaon
  condensation in hyperon-mixed matter}, Phys. Lett. B \textbf{820}, 136587
  (2021), \texttt{2106.03449}. \doiwoc{10.1016/j.physletb.2021.136587}

\bibitem{Lutz:1997wt}
M.~Lutz, {Nuclear kaon dynamics}, Phys. Lett. \textbf{B426}, 12 (1998),
  \texttt{nucl-th/9709073}. \doiwoc{10.1016/S0370-2693(98)00299-8}

\bibitem{Ramos:1999ku}
A.~Ramos, E.~Oset, {The Properties of $\bar K$ in the nuclear medium}, Nucl.
  Phys. \textbf{A671}, 481 (2000), \texttt{nucl-th/9906016}.
  \doiwoc{10.1016/S0375-9474(99)00846-5}

\bibitem{Tolos:2000fj}
L.~Tolos, A.~Ramos, A.~Polls, T.T.S. Kuo, {Partial wave contributions to the
  anti-kaon potential at finite momentum}, Nucl. Phys. \textbf{A690}, 547
  (2001), \texttt{nucl-th/0007042}. \doiwoc{10.1016/S0375-9474(00)00711-9}

\bibitem{Tolos:2002ud}
L.~Tolos, A.~Ramos, A.~Polls, {The Anti-kaon nuclear potential in hot and dense
  matter}, Phys. Rev. \textbf{C65}, 054907 (2002), \texttt{nucl-th/0202057}.
  \doiwoc{10.1103/PhysRevC.65.054907}

\bibitem{Tolos:2006ny}
L.~Tolos, A.~Ramos, E.~Oset, {Chiral approach to antikaon s and p-wave
  interactions in dense nuclear matter}, Phys. Rev. \textbf{C74}, 015203
  (2006), \texttt{nucl-th/0603033}. \doiwoc{10.1103/PhysRevC.74.015203}

\bibitem{Tolos:2008di}
L.~Tolos, D.~Cabrera, A.~Ramos, {Strange mesons in nuclear matter at finite
  temperature}, Phys. Rev. \textbf{C78}, 045205 (2008), \texttt{0807.2947}.
  \doiwoc{10.1103/PhysRevC.78.045205}

\bibitem{Cabrera:2014lca}
D.~Cabrera, L.~Tolos, J.~Aichelin, E.~Bratkovskaya, {Antistrange meson-baryon
  interaction in hot and dense nuclear matter}, Phys. Rev. \textbf{C90}, 055207
  (2014), \texttt{1406.2570}. \doiwoc{10.1103/PhysRevC.90.055207}

\bibitem{Lutz:2007bh}
M.F.M. Lutz, C.L. Korpa, M.~Moller, {Antikaons and hyperons in nuclear matter
  with saturation}, Nucl. Phys. \textbf{A808}, 124 (2008), \texttt{0707.1283}.
  \doiwoc{10.1016/j.nuclphysa.2008.05.008}

\bibitem{Blacker:2023opp}
S.~Blacker, H.~Kochankovski, A.~Bauswein, A.~Ramos, L.~Tolos, {Thermal behavior
  as indicator for hyperons in binary neutron star merger remnants}, Phys. Rev.
  D \textbf{109}, 043015 (2024), \texttt{2307.03710}.
  \doiwoc{10.1103/PhysRevD.109.043015}

\bibitem{Kochankovski:2022rid}
H.~Kochankovski, A.~Ramos, L.~Tolos, {Equation~of state for hot hyperonic
  neutron star matter}, Mon. Not. Roy. Astron. Soc. \textbf{517}, 507 (2022),
  [Erratum: Mon.Not.Roy.Astron.Soc. 518, 6376], \texttt{2206.11266}.
  \doiwoc{10.1093/mnras/stac2671}

\bibitem{Raduta:2022elz}
A.R. Raduta, {Equations of state for hot neutron stars-II. The role of exotic
  particle degrees of freedom}, Eur. Phys. J. A \textbf{58}, 115 (2022),
  \texttt{2205.03177}. \doiwoc{10.1140/epja/s10050-022-00772-0}

\end{thebibliography}

\end{document}